\documentclass[10pt,a4paper]{article}
\usepackage[utf8]{inputenc}
\usepackage[T1]{fontenc}
\usepackage{amsmath}
\usepackage{amssymb}
\usepackage{makeidx}
\usepackage{graphicx}
\usepackage{caption}
\usepackage{subcaption}
\usepackage{amsmath}
\usepackage{mathtools}
\usepackage{amsfonts}

\begin{document}

\title{Tunneling probability for the birth of an universe with radiation in Ho\v{r}ava-Lifshitz theory}

\author{G. Oliveira-Neto and A. Oliveira Castro Júnior\\
Departamento de F\'{\i}sica, \\
Instituto de Ci\^{e}ncias Exatas, \\ 
Universidade Federal de Juiz de Fora,\\
CEP 36036-330 - Juiz de Fora, MG, Brazil.\\
gilneto@fisica.ufjf.br, alessandroocj@protonmail.com
\and G. A. Monerat\\
Departamento de Modelagem Computacional, \\
Instituto Polit\'{e}cnico, \\
Universidade do Estado do Rio de Janeiro, \\
CEP 28.625-570, Nova Friburgo - RJ - Brazil.\\
monerat@uerj.br}

\maketitle

\begin{abstract}
In the present work, we study the birth of a homogeneous and isotropic Friedmann–Lemaître–Robertson–Walker (FLRW) cosmological model, 
considering Ho\v{r}ava-Lifshitz (HL) as the gravitational theory. The matter content of the model is a radiation perfect fluid. In order to
study the birth of the universe in the present model, we consider the quantum cosmology mechanism of {\it creation from nothing}. In that mechanism, the universe
appears after the wavefunction associated to that universe tunnels through a potential barrier. We started studying the classical model.
We draw the phase portrait of the model and identify qualitatively all types of dynamical behaviors associated to it. Then, we write
the Hamiltonian of the model and apply the Dirac's quantization procedure to quantize a constrained theory. We find the appropriate 
Wheeler-DeWitt equation and solve it using the Wentzel–Kramers–Brillouin (WKB) approximation. Using the WKB solution, to the Wheeler-DeWitt
equation, we compute the tunneling probabilities for the birth of that universe ($TP_{WKB}$). Since the WKB wavefunction depends on the radiation 
energy ($E$) and the free parameters coming from the HL theory ($g_c$, $g_r$, $g_s$, $g_\Lambda$), we compute the behavior of $TP_{WKB}$ as a 
function of $E$ and all the HL's parameters $g_c$, $g_r$, $g_s$, $g_\Lambda$.
\end{abstract}



\section{Introduction}
\label{intro}

Some years ago Petr Ho\v{r}ava introduced a new geometrical theory of gravity \cite{horava}. His main motivation
was to obtain a renormalizable and unitary theory of gravity. One important ingredient of that theory, nowadays 
known as Ho\v{r}ava-Lifshitz theory, is an anisotropic scaling between space and time. That anisotropy is given 
by a critical exponent $z$. Whenever that anisotropy or asymmetry between space and time is important the Lorentz symmetry 
is broken. It happens at high energies. At low energies HL theory tends to General Relativity (GR), which implies
$z=1$, and the Lorentz symmetry is recovered. In his original paper, Ho\v{r}ava formulated his theory using some
elements of the Arnowitt-Deser-Misner (ADM) formalism \cite{misner}. In that formalism the four dimensional spacetime
metric ($g_{\alpha \beta}$) is decomposed in terms of three dimensional metrics of spacelike sections ($h_{i j}$), the 
shift vector ($N_i$) and the lapse function ($N$). In the most general situation all these quantities should depend on
space and time. In order to facilitate the work with his theory, Ho\v{r}ava considered two simplifications. In the
first one, he imposed that $N$ should depend only on the time variable \cite{horava}. This assumption is known as 
the {\it projectable condition}. In the second one, he reduced the number of terms contributing to the potential component 
of his theory \cite{horava}. It is known as the {\it detailed balance condition}. Unfortunately, it was shown that the projectable 
version of the HL theory, with the detailed balance condition, have massive ghosts and instabilities \cite{mattvisser1,wang}.
Since its introduction many physicists have investigated different aspects of HL theory. In particular, when applied to cosmology
it produced very interesting results \cite{bertolami,saa,kord,pedram2,misonoh,ivano,bramberger,gil3,gil,gil2,gao,leon,cordero,nilsson,gao1,compean,gao2,tavakoli,vicente,gao3,compean1,compean2,czuchry,nilsson1}. 
Some important aspects of the HL theory are discussed in Ref.\cite{wang1}.

The main motivation in order to quantize GR was the removal of physical singularities present in many cosmological as well as 
astrophysical solutions to Einstein's equations. The first systematic approach toward the quantization of that theory was 
the canonical quantization. It started with the important works from P. A. M. Dirac on the correct Hamiltonian formulation of GR 
\cite{dirac,dirac1,dirac2,dirac3}. After that, ADM gave an influential geometrical interpretation of the Hamiltonian formulation
of GR \cite{misner}. Next, the canonical quantization of GR was performed, following Dirac's instructions in order to quantize a 
constrained theory \cite{dirac,dirac4}, which lead to the famous Wheeler-Dewitt equations \cite{dewitt,wheeler}. When
one applies that formalism to cosmological models one obtains a theory called Quantum Cosmology (QC). For a detailed introduction to
QC see \cite{halliwell,paulo,kiefer,julio}. One of the main applications
of QC is the description of the beginning of the Universe. GR tells us that our Universe started in a huge explosion called 
{\it Big Bang}, which created space, time and some of the matter present in our Universe. GR, also, tell us that the {\it Big Bang} 
is a spacetime singularity. It means that GR loses its predictive power in the spacetime location of that catastrophic event. One of
the most important results of QC is to offer a possible solution to the singular beginning of our Universe. That solution coming from
QC uses, as a fundamental ingredient, the quantum mechanical tunneling mechanism and is usually referred to as {\it spontaneous creation
of the Universe from nothing} (SCUFN)\cite{grishchuk, vilenkin, vilenkin1, vilenkin2, hawking, linde, rubakov, vilenkin3}. In the SCUFN
the Universe is at the origin of time and space and confined by a potential barrier. Since the Universe is a quantum mechanical system it
may tunnel through the potential barrier and emerge to the right of it with a finite size. That moment marks the beginning of the Universe.
Since the Universe starts with a finite size, it is free from the {\it Big Bang} singularity. In the literature, there are some 
works using these ideas in order to describe the birth of the universe in different cosmological models \cite{paulo1,acacio,germano,germano1,germano2,rocha,gil4}.

In the present paper, we want to study the birth of the universe in a cosmological model where the gravitational sector is described by the HL theory.
Therefore, we consider a homogeneous and isotropic FLRW spacetime, with positively curved spatial sections ($k=1$).
The matter content of the model is a radiation perfect fluid. With respect to the HL theory, we consider the projectable version of that
theory without the detailed balance condition. We start by writing the total Hamiltonian of the gravitational sector of the model, with the
aid of the ADM formalism. Then, we obtain the total Hamiltonian of the matter sector of the model, with the aid of the Schutz variational
formalism \cite{schutz,schutz1}. Adding both total Hamiltonians, we find the total Hamiltonian of the model. From that total Hamiltonian, we
notice that the potential barrier may have two different shapes. In the first one, it has a single barrier shape. In the second one, it has a 
well followed by a barrier shape. Due to the fact that the first barrier shape is a particular case of the second one, we restrict our attention 
to study the birth of the universe only for the second barrier shape. In order to learn more about that particular model, we study the classical 
dynamics of the scale factor. First, using the total Hamiltonian of the model we draw a phase portrait of the model, which gives a qualitative 
idea on the possible scale factor behaviors. Then, using once more the total Hamiltonian of the model we find and solve the Hamilton's equations. 
We quantitatively obtain all possible scale factor solutions for the model. After that, we quantize the model following the Dirac's
formalism for constrained theories \cite{dirac,dirac4}, which produces the Wheeler-DeWitt equation of the model. We solve that
equation using the WKB approximation and compute the appropriate tunneling probabilities for the birth of the universe ($TP_{WKB}$). Since $TP_{WKB}$ is a
function of the radiation fluid energy ($E$) and all the parameters coming from the HL theory ($g_\Lambda$, $g_c$, $g_s$, $g_r$), we investigate, 
separately, how it depends on all these quantities.

In Section \ref{classical}, we derive the total Hamiltonian, draw a phase portrait and solve the Hamilton's equations for a homogeneous and
isotropic HL cosmological model, with constant positive spatial sections and coupled to a radiation perfect fluid. In Section 
\ref{quantization}, we quantize the model and solve the resulting Wheeler-DeWitt equation, using the WKB approximation.  
In Section \ref{results}, we compute the WKB tunneling probabilities for the universe to tunnel through the potential barrier.
We investigate how $TP_{WKB}$ depends on $g_\Lambda$, $g_c$, $g_s$, $g_r$ and $E$. Finally, in Section \ref{CC}, we summarize the
main points and results of the paper. In Appendix \ref{SF}, we compute in details, with the aid of the Schutz variational formalism, the total Hamiltonian for the radiation fluid.

\section{The Classical Model}
\label{classical}

We start by considering an isotropic and homogeneous Universe described by the FLRW metric,
\begin{equation}
    ds^2 = -N^2(t)dt^2 + a^2(t) \left[ \frac{dr^2}{1-kr^2} + r^2 (d\theta^2 + sin^2 \theta d\phi^2) \right],
\label{metric}
\end{equation}
where $N(t)$ is called the \textit{lapse function} and $k$ provides the curvature of the spacelike sections. $k$ may take the values $k=-1,0,+1$, which gives, respectively, open, flat, or closed geometries. In the present work, we consider that the constant curvatures of the spacelike sections are positive ($k=+1$). We use the natural unit system where: $\hbar = 8\pi G = c = k_{B} = 1$. The matter content of the model is given by a radiation perfect fluid having a four-velocity given by: $U^\mu = \delta_0^\mu$, in the co-moving coordinate system. The energy-momentum tensor is written as,
\begin{equation}
    T_{\mu\nu} = (\rho + p) U_\mu U_\nu -pg_{\mu\nu},
\end{equation}
where $p$ and $\rho$ are the pressure and energy density of the fluid, respectively, and the Greek indexes $\mu$ and $\nu$ run from zero to three. Furthermore, the equation of state for the radiation perfect fluid is,
\begin{equation}
\label{eqstate}
p = \rho/3.
\end{equation}
Selecting the projectable version of HL theory in the absence of the detailed balance condition, for $z=3$ in $3+1$-dimensions, the action is given by,
\begin{equation}
\begin{split}
    S_{HL} &= \frac{M_p^2}{2} \int d^3x dt N \sqrt{h} [ K_{ij}K^{ij} - \lambda K^2 - g_0 M_{p}^{2} - g_1 R - M_{p}^{-2}(g_2 R^2 + g_3R_{ij}R^{ij}) \\
    & - M_{p}^{-4}(g_4 R^3 + g_5 R R^{i}_{j}R^{j}_{i} + g_6 R^{i}_{j} R^{j}_{k}R^{k}_{i} + g_7 R \nabla^2 R + g_8 \nabla_i R_{jk} \nabla^i R^{jk}) ], 
\end{split}
\label{action}
\end{equation}
where the $g_l$'s ($l = 0,..,8$) and $\lambda$ are parameters linked to HL theory, $M_p$ is the Planck mass, $K_{ij}$ are the components of the extrinsic curvature tensor which has its trace denoted by $K$. Also, $R_{ij}$ are the components of the Ricci tensor computed for the metric of the three dimensional spatial sections, namely $h_{ij}$, and  $\nabla_i$ represents the covariant derivative in the $i$ direction. The Latin indices $i$, $j$ and $k$ run from one to three. It is also important to comment that General Relativity is retrieved once one takes the limit $\lambda \to 1$.

Introducing the metric of the spatial sections that comes from the FLRW spacetime Eq. (\ref{metric}) in the action Eq. (\ref{action}) and adopting the following values: $g_0 M_p^2 = 2\Lambda$ and $g_1 = -1$, we manage to write,
\begin{equation}
\begin{split}
    S_{HL} &= \xi \int dt N \{ -\frac{\Dot{a}^2 a}{N^2} + \frac{1}{3(3\lambda - 1)} ( 6ka - 2\Lambda a^3 - \frac{12k^2}{a M_{p}^{2}}(3g_2 + g_3) \\
    & - \frac{24k^3}{a^3 M_{p}^{4}}(9g_4 + 3g_5 + g_6) )  \} .
\end{split}
\label{complete_action}
\end{equation}
Here, one has that,  
\begin{equation}
    \xi = \frac{3M_p^2(3\lambda-1)}{2}\int d^3x \frac{r^2sin\theta}{\sqrt{1-kr^2}}.
\end{equation}
As a matter of simplicity let us consider $\xi =1$. Using this condition, $S_{HL}$ Eq. (\ref{complete_action}) may be written as,
\begin{equation}
    S_{HL} = \int dt N \left[ -\frac{\Dot{a}^2 a}{N^2} + g_c k a - g_\Lambda a^3 - g_r \frac{k^2}{a} - g_s \frac{k^3}{a^3} \right].
\end{equation}
where,
\begin{equation}
    g_c = \frac{2}{3\lambda -1},
\label{first}
\end{equation}
\begin{equation}
    g_\Lambda = \frac{2\Lambda}{3(3\lambda-1)},
\end{equation}
\begin{equation}
    g_r = \frac{4}{(3\lambda-1)M_p^2} (3g_2+g_3),
\end{equation}
\begin{equation}
    g_s = \frac{8}{(3\lambda-1)M_p^4}(9g_4 + 3g_5 + g_6).
\label{last}
\end{equation}
The equations running from (\ref{first}-\ref{last}) provide us with $g_c, g_\Lambda, g_r$, and $g_s$, the new coupling constants. 
It is important to mention that $g_c$ must be greater than zero ($g_c > 0$), because it is associated to the curvature coupling 
constant. The others may take positive or negative values. 
One may recall that $S = \int L dt$ so, for this action, one gets,
\begin{equation}
    \mathcal{L}_{HL} = N \left[ -\frac{\Dot{a}^2 a}{N^2} + g_c k a - g_\Lambda a^3 - g_r \frac{k^2}{a} - g_s \frac{k^3}{a^3} \right].
\label{HLLagrangian}
\end{equation}
This allows us to compute the canonical momentum
\begin{equation}
    p_a = \frac{\partial \mathcal{L}_{HL}}{\partial \Dot{a}} = \frac{\partial}{\partial \Dot{a}} \left[ -\frac{\Dot{a}^2 a}{N}\right] = -\frac{2 \Dot{a} a}{N}
\label{momentum}
\end{equation}
Introducing the momentum Eq. (\ref{momentum}) and the Lagrangian Eq. (\ref{HLLagrangian}) into the total Hamiltonian definition and fixing $k=+1$, as mentioned above, one obtains,
\begin{equation}
\label{HLTotalHamiltonian}
     N\mathcal{H}_{HL} = N \left[ -\frac{p_a^2}{4a} - g_c a + g_\Lambda a^3 + \frac{g_r}{a} + \frac{g_s}{a^3} \right],
\end{equation}
which is the total Hamiltonian for the HL sector.
The total Hamiltonian for the matter sector ($N\mathcal{H}_{f}$), described by a radiation perfect fluid, is obtained through the Schutz Formalism. In Appendix \ref{SF}, we compute in details $N\mathcal{H}_{f}$, whose expression is given by Eq. (\ref{hamiltoniana fluido_2}).
Adding it to the HL total Hamiltonian (\ref{HLTotalHamiltonian}), one finds,
\begin{equation}
    N\mathcal{H}_{T} = N \left[ -\frac{p_a^2}{4a} - g_c a + g_\Lambda a^3 + \frac{g_r}{a} + \frac{g_s}{a^3} + \frac{p_T}{a}\right].
\label{hamiltonian}
\end{equation}
In order to learn more about the scale factor classical behavior one should find and solve the Hamilton’s equations. These equations are computed with the aid of $N\mathcal{H}_{T}$ Eq. (\ref{hamiltonian}) and they are,
\begin{equation}
\centering
\begin{cases}
\Dot{a} &= \frac{\partial N\mathcal{H}_{T}}{\partial p_a} = -\frac{p_a}{2}, \\

 \Dot{p_a} &= -\frac{\partial N\mathcal{H}_{T}}{\partial a} = -N \left( \frac{p_a^2}{4a^2} - g_c + 3 g_\Lambda a^2 - \frac{g_r}{a^2} - 3 \frac{g_s}{a^4} - \frac{p_T}{a^{2}}\right), \\

 \Dot{T} &= \frac{\partial N\mathcal{H}_{T}}{\partial p_T} = 1, \\

 \Dot{p_T} &= \frac{\partial N\mathcal{H}_{T}}{\partial T} = 0, \\
\end{cases}
\label{HamiltonEq}
\end{equation}
\noindent where the gauge $N=a$ was used and the dot means derivative with respect to the conformal time $d\eta \equiv Ndt$. If one imposes the constraint equation $\mathcal{H}_T = 0$, and rewrite the resulting equation in terms of $\dot{a}$, with the aid of the first Hamilton's equation Eq. (\ref{HamiltonEq}), one finds,
\begin{equation}
\label{friedmanneq}
    \dot{a}^2 + V_c(a) = 0,
\end{equation}
where the classical potential $V_c(a)$ is given by,
\begin{equation}
\label{classicalpotential}
    V_c(a) = g_c a^2 - g_\Lambda a^4 - g_r - \frac{g_s}{a^2} - p_T.
\end{equation}

\begin{figure}[!tbp]
  \centering
  \subfloat[]{\includegraphics[width=0.45\textwidth]{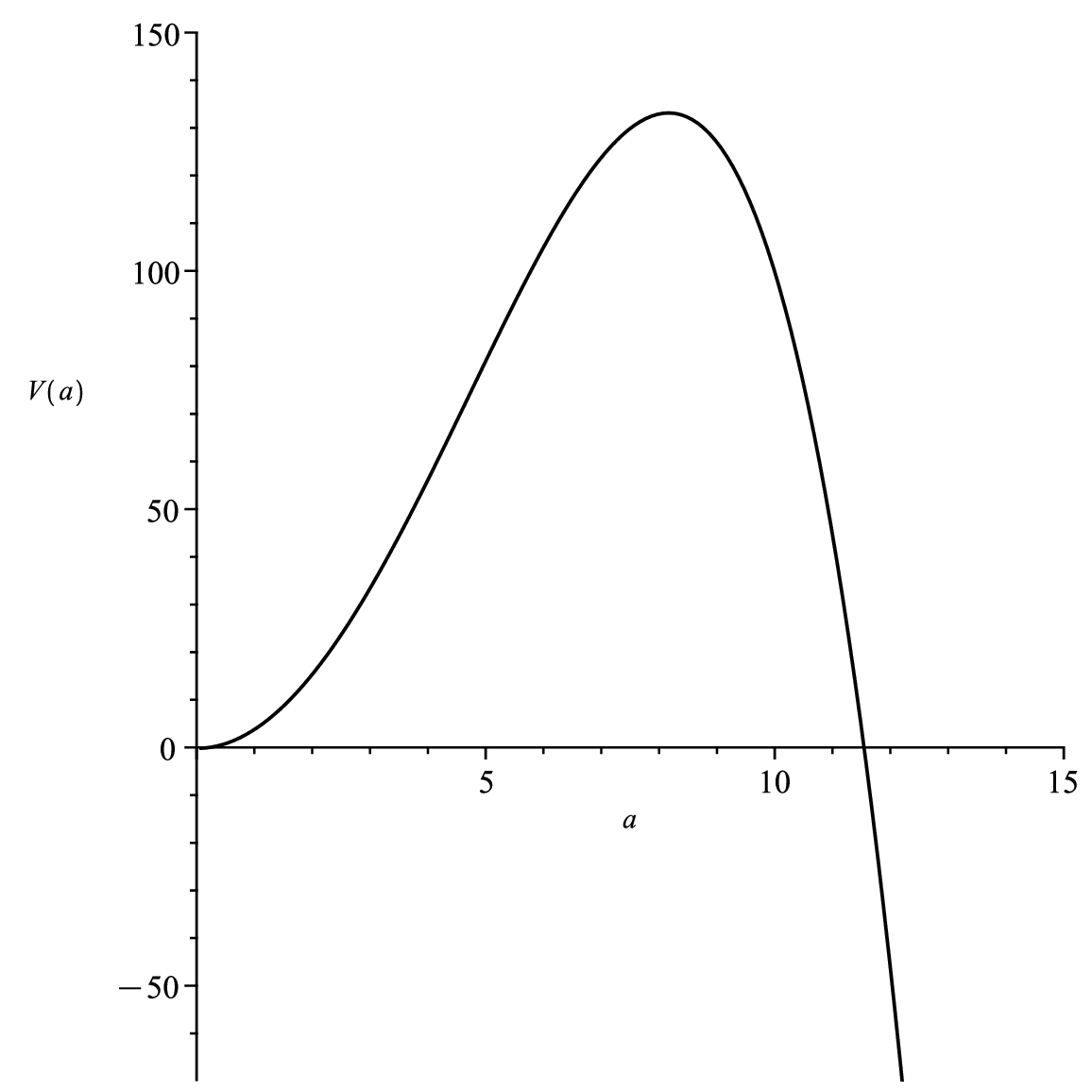}\label{potential-1}}
  \hfill
  \subfloat[]{\includegraphics[width=0.45\textwidth]{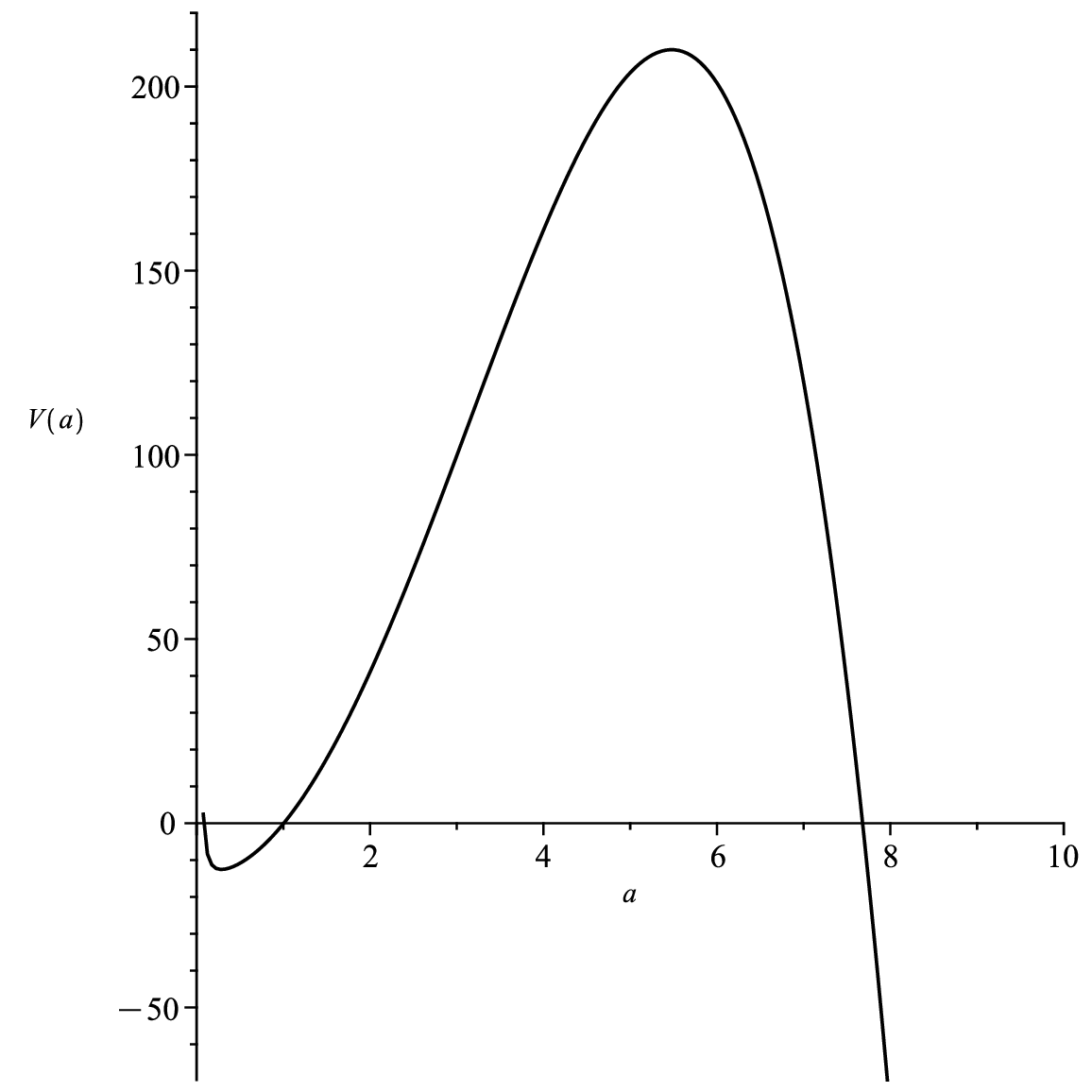}\label{potential-2}}
  \caption{Classical potential $V_c(a)$ behavior for two different situations. In situation (a) one has $g_c = 4.0, g_\Lambda = 0.03, g_r = 0.2$ and $g_s = -0.001$ whereas in situation (b) one has $g_c = 15.0, g_\Lambda = 0.25, g_r = 15.0$ and $g_s = -0.1$. For both situations $p_T = 0$}
\label{figpotential}
\end{figure}

\begin{figure}[!tbp]
\includegraphics[width=0.45\textwidth]{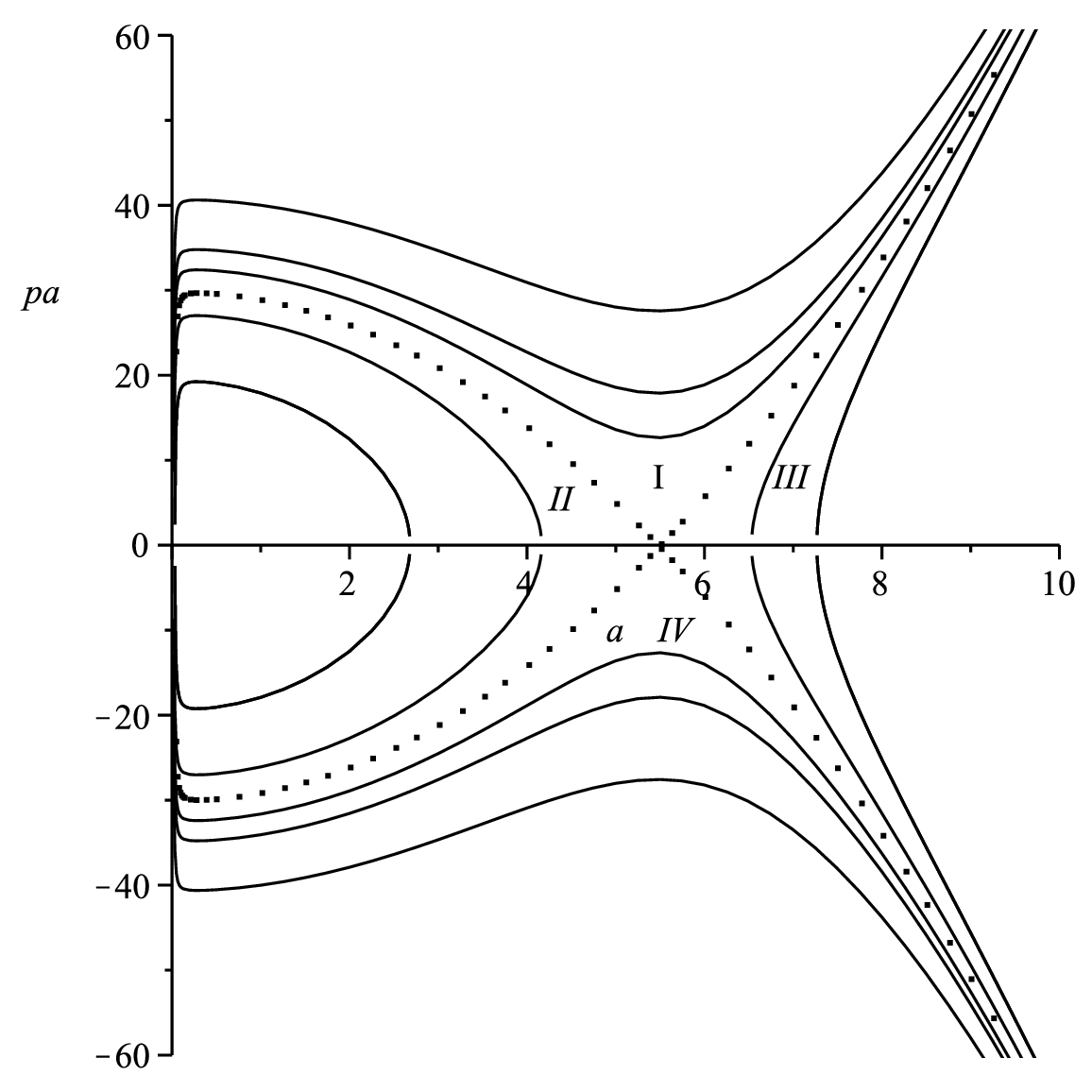}
  \caption{Phase portrait for the model where the potential has the shape of a well followed by a barrier. It uses the same values of the HL's parameters as in Figure (\ref{potential-2}). On the other hand, $p_T$ varies from $0$ to $400$. The dotted lines are called separatrices because they separate different types of solutions.}
\label{figphaseportrait}
\end{figure}

As mentioned above $g_c > 0$. $p_T$ is also positive because it is associated to the fluid energy density. 
In order to obtain well defined barrier type potentials, we study, here, models in which $g_r \geq 0$, $g_\Lambda > 0$ and $g_s < 0$. For the choice $g_s < 0$, the potential develops an infinity positive barrier at the origin and those models are free from the big bang singularity.
Taking in account those choices in the values of the HL's parameters, we notice that
$V_c(a)$ Eq. (\ref{classicalpotential}) may have two different shapes depending on the values of these parameters. The first one has a single barrier and the second one has a well followed by a barrier. Examples for each potential shape are shown in Figures (\ref{potential-1}) and (\ref{potential-2}), respectively. The potential shown in Figure (\ref{potential-1}) has a global maximum of $V_{max_1} = 133.1333483$ while the one presented in Figure (\ref{potential-2}) has a global maximum of $V_{max_2} = 210.0033333$ and a local minimum of $V_{min_2} = -12.55217920$. Since the first type of potential (barrier) is much simpler and may be considered as a particular case of the second one (well followed by a barrier), in the present paper, we restrict our attention to study the second type of potential.
In order to learn more about the present model, we study the classical dynamics of the scale factor. We start our study by drawing a phase
portrait of the model. Figure (\ref{figphaseportrait}) shows a phase portrait of the well followed by a barrier type of potential. There, the values of the HL's parameters are the same ones used in Figure (\ref{potential-2}).

\begin{figure}[!tbp]
  \centering
  \subfloat[]{\includegraphics[width=0.45\textwidth]{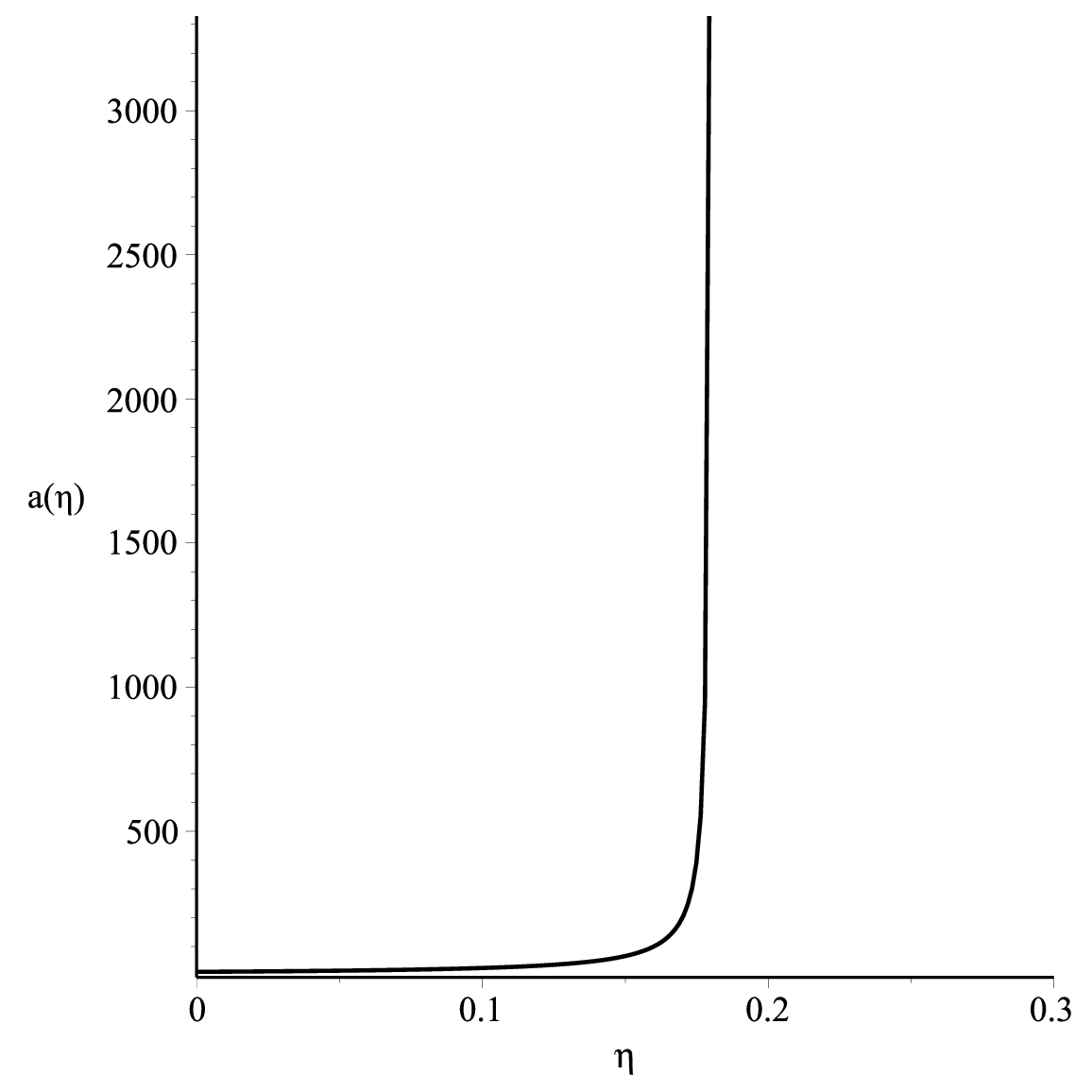}\label{Fig_Sol_Exp}}
  \hfill
  \subfloat[]{\includegraphics[width=0.45\textwidth]{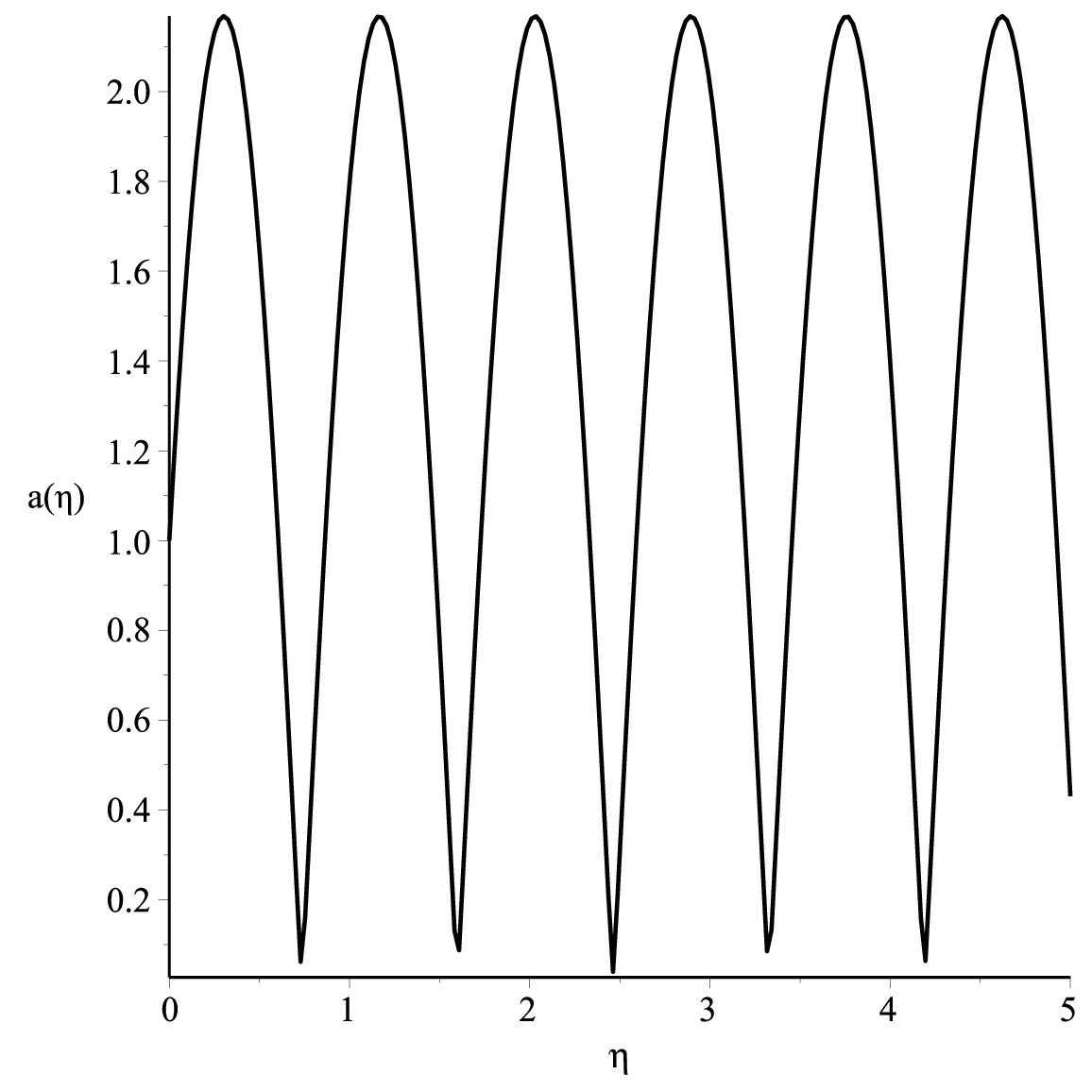}\label{Fig_Sol_Per}}
  \caption{Classical solutions $a(\eta) \times \eta$ for the cases of expansion (a) and periodic movement (b).}
\label{figsolutions1}
\end{figure}

\begin{figure}[!tbp]
  \centering
  \subfloat[]{\includegraphics[width=0.45\textwidth]{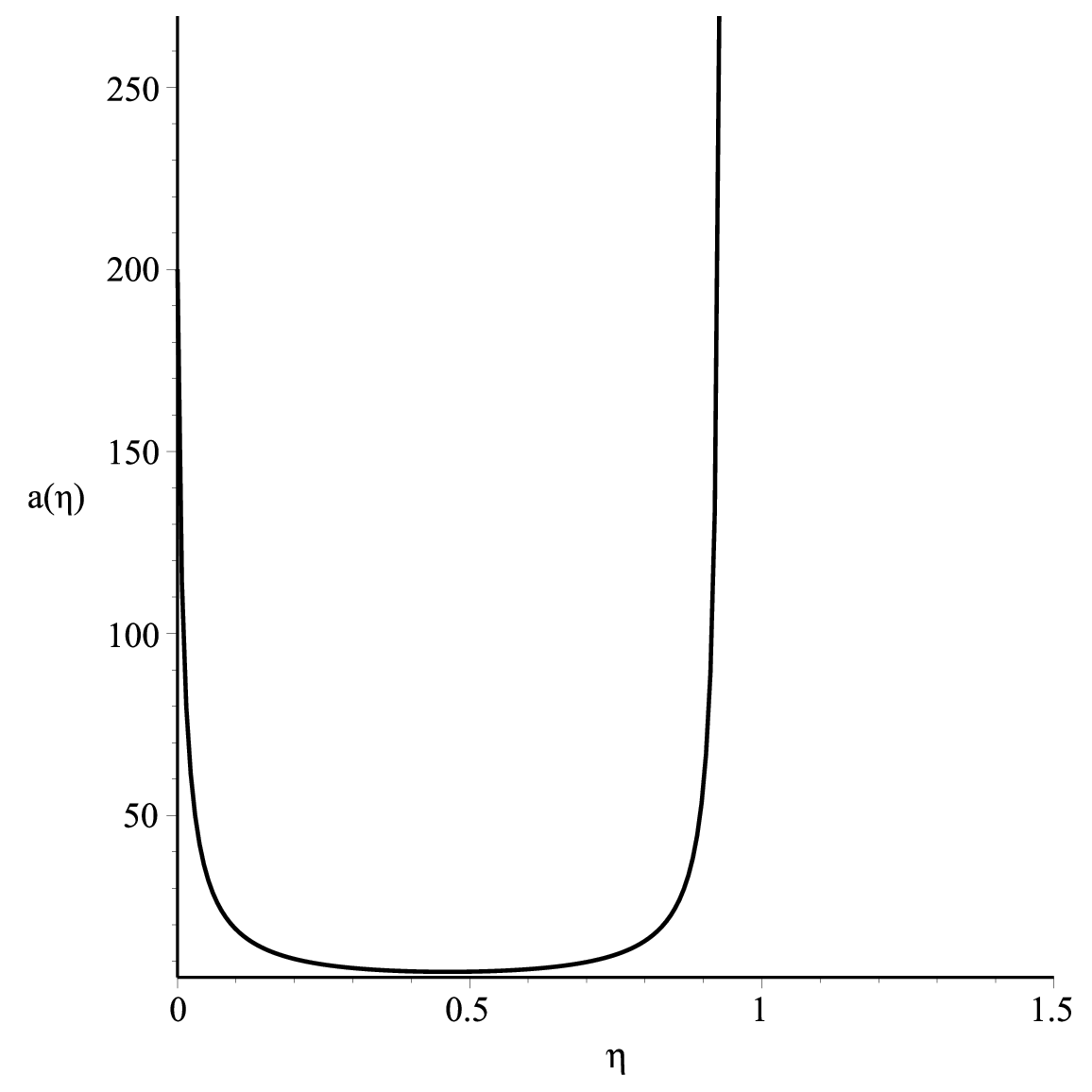}\label{Fig_Sol_Boun}}
  \hfill
  \subfloat[]{\includegraphics[width=0.45\textwidth]{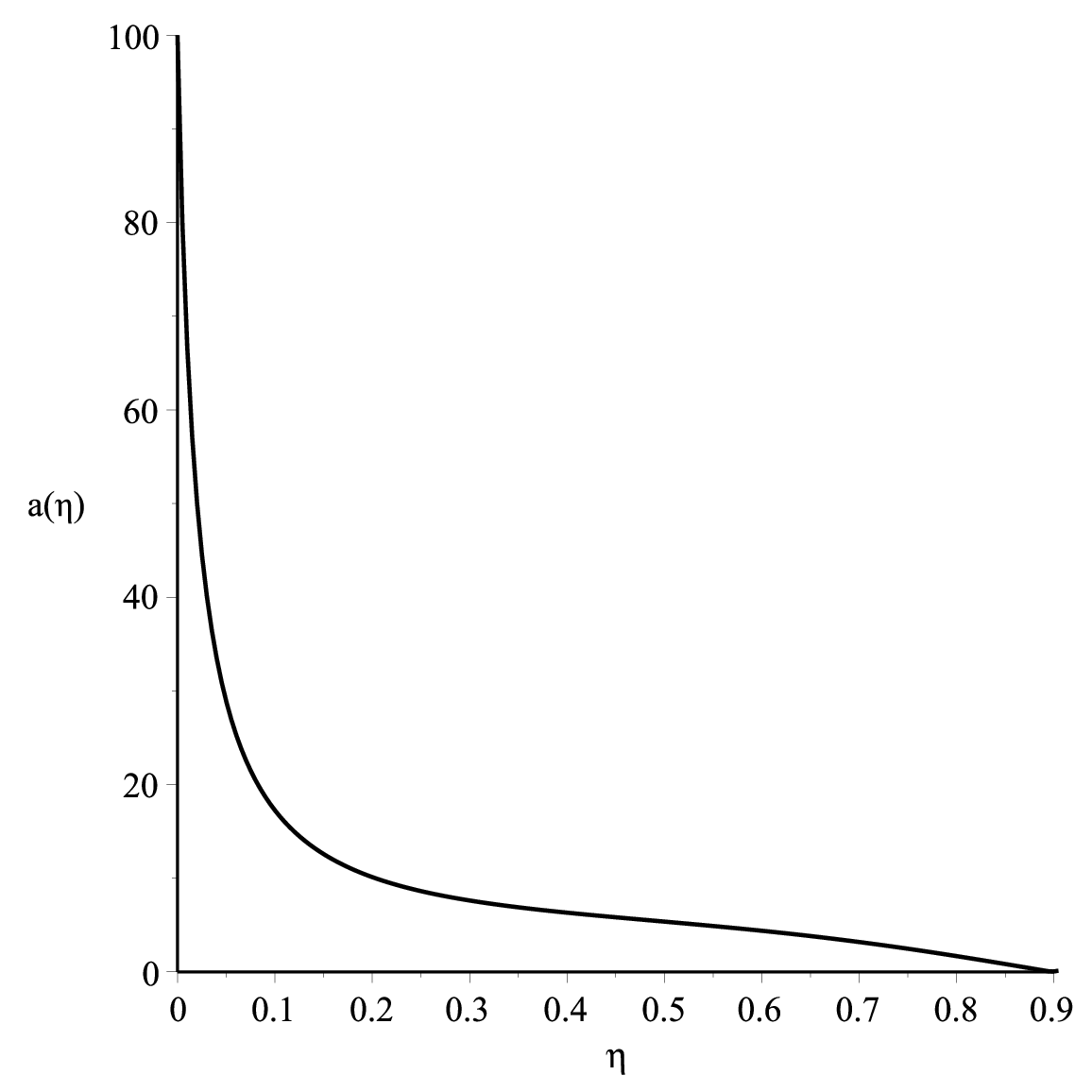}\label{Fig_Sol_Cont}}
  \caption{Classical solutions $a(\eta) \times \eta$ for the cases of bouncing (a) and contraction (b).}
\label{figsolutions2}
\end{figure}

After combining some of the Hamilton's equations (\ref{HamiltonEq}), we obtain the following second-order differential equation for the scale factor,
\begin{equation}
\label{classicaleq}
    \frac{\partial^2 a(\eta)}{\partial \eta^2} + g_c a(\eta) - 2g_\Lambda a(\eta)^3 + \frac{g_s}{a(\eta)^3} = 0.
\end{equation}
Solving Eq. (\ref{classicaleq}) using the appropriate initial conditions, we find four different types of classical solutions for $V_c(a)$
Eq. (\ref{classicalpotential}), when it has the shape of a well followed by a barrier Figure \ref{potential-2}. The four types of solutions may be identified from the phase portrait Figure \ref{figphaseportrait}. They are: (i) An expansive solution where the scale factor starts expanding from a small value and continue expanding, in an accelerated rate, to infinity. This type of solution appears in region I of Figure \ref{figphaseportrait}. An example is given in Figure \ref{Fig_Sol_Exp}; (ii) A periodic solution where the scale factor starts expanding from a small value, reaches a maximum and then contracts back to the initial small value. This type of solution appears in region II of Figure \ref{figphaseportrait}. An example is given in Figure \ref{Fig_Sol_Per}; (iii) A bouncing solution where the scale factor starts contracting from a large value, reaches a minimum value and then expands, in an accelerated rate, to infinity. This type of solution appears in region III of Figure \ref{figphaseportrait}. An example is given in Figure \ref{Fig_Sol_Boun}; (iv) A solution where the scale factor starts contracting from a large value and continue contracting to a small value. This type of solution appears in region IV of Figure \ref{figphaseportrait}. An example is given in Figure \ref{Fig_Sol_Cont}.

\section{Canonical Quantization}
\label{quantization}

In the present section, we want to quantize the model according to Dirac's formalism for quantization of constrained systems. It starts by replacing the canonical variables $a$ and $T$ and theirs conjugated momenta by operators. Then, one must introduce the wavefunction of the Universe which is a function of the operators associated to the canonical variables, $\Psi(a,T)$. The canonically conjugated momenta to $a$ and $T$, respectively $p_a$ and $p_T$, have the following operators expressions,
\begin{equation}
\label{operators}
    p_a \longrightarrow -i\frac{\partial}{\partial a},\ \ \ p_T \longrightarrow -i\frac{\partial}{\partial T}.
\end{equation}
Next, one must impose the constraint equation, 
\begin{equation}
\label{constraint}
\hat{H}\Psi(a,T) = 0,
\end{equation}
where $\hat{H}$ is obtained by replacing the operators expressions of $a$, $T$, $p_a$ and $p_T$ in the total Hamiltonian Eq. (\ref{hamiltonian}). The constraint equation (\ref{constraint}), gives rise to the so-called Wheeler-DeWitt equation. Using the total Hamiltonian Eq. (\ref{hamiltonian}), the operators expressions of $p_a$ and $p_T$ Eq. (\ref{operators}), the gauge $N=a$ and the transformation $\tau = -T$, one obtains,
\begin{equation}
    \left( \frac{1}{4}\frac{\partial^2}{\partial a^2} - g_c a^2 + g_\Lambda a^4 + g_r + \frac{g_s}{a^2} \right)\Psi = -i \frac{\partial \Psi}{\partial \tau}.
\label{WdW}
\end{equation}

Presuming that one may write the solution to equation (\ref{WdW}) as $\Psi(a,\tau) = \psi(a)e^{-iE\tau}$, after some calculations, one finds, 
\begin{equation}
    \frac{d^2\psi(a)}{da^2} + 4 \left( E - V(a) \right)\psi(a) = 0,
\label{WdW-2}
\end{equation}
where $V(a)$ is given by,
\begin{equation}
\label{potential}
V(a) = g_c a^2 - g_\Lambda a^4 - g_r - \frac{g_s}{a^2}.
\end{equation}

Now, suppose that the WKB approximation is valid for the present model \cite{merzbacher,griffiths}. Once obtained, the WKB solution will be applied to compute the tunneling probabilities through the potential barrier $V(a)$. 
Following the WKB approximation, we write $\psi(a)$ in Eq. (\ref{WdW-2}) as $\psi(a)=\mathcal{A}(a)e^{i\varphi(a)}$. Introducing this ansatz in Eq. (\ref{WdW-2}) and considering that $d^2\mathcal{A}(a)/da^2$ is insignificant compared to the other terms in the resulting equation, we may write the following solution to that equation,
\begin{equation}
\psi(a)=\begin{cases}
          \frac{A}{\sqrt{K(a)}}exp\left( i \int_{a}^{a_l}K(a)da \right) + \frac{B}{\sqrt{K(a)}}exp\left( -i \int_{a}^{a_l}K(a)da \right) \quad & \, 0\leq a\leq a_l, \\
          
          \frac{C}{\sqrt{k(a)}}exp\left( - \int_{a_l}^{a}k(a)da \right) + \frac{D}{\sqrt{k(a)}}exp\left( \int_{a_l}^{a}k(a)da \right) \quad &\, a_l\leq a \leq a_r, \\
          
          \frac{F}{\sqrt{K(a)}}exp\left( i \int_{a_r}^{a}K(a)da \right) + \frac{G}{\sqrt{K(a)}}exp\left( -i \int_{a_r}^{a}K(a)da \right) \quad & \, a_r \leq a < \infty, \\
     \end{cases}
\end{equation}
where,
\begin{equation}
\label{ks}
\begin{cases}
          K(a) = 2 \sqrt{\left(E - V(a) \right)} \quad & \, E > V, \\
          
          k(a) = 2 \sqrt{\left( V(a) - E \right)} \quad & \, E < V, \\
\end{cases}
\end{equation}
$a_l$ is the value of $a$ where the energy $E$ intercepts, at the left, the potential $V(a)$ (\ref{potential}), $a_r$ is the value of 
$a$ where the energy $E$ intercept, at the right, the potential $V(a)$ (\ref{potential}) and $A$, $B$, $C$, $D$, $F$ and $G$ are constants.

Using matrix notation, one can relate A and B to F and G by, 
\begin{equation}
\begin{pmatrix}
A \\
B 
\end{pmatrix}  =  \begin{pmatrix}
2\theta + \frac{1}{2\theta} & i\left(2\theta - \frac{1}{2\theta} \right) \\
-i\left(2\theta - \frac{1}{2\theta} \right) &  2\theta + \frac{1}{2\theta}
\end{pmatrix}
\begin{pmatrix}
F \\
G 
\end{pmatrix},
\label{relationABFG}
\end{equation}
where one has the parameter $\theta$, which gives the barrier height and length, given by,
\begin{equation}
\label{theta0}
    \theta = exp\left( \int_{a_l}^{a_r} k(a)da \right).
\end{equation}

Now that the WKB solution was obtained, one has to compute the transmission coefficient, or tunneling probability, denoted by $TP_{WKB}$, through the potential barrier. If one considers that there is no wave coming from the right ($G=0$), 
\begin{equation}
    TP_{WKB} = \frac{|F|^2}{|A|^2}.
\end{equation}
\noindent From equation (\ref{relationABFG}), one has,
\begin{equation}
    A = \frac{1}{2}\left(2\theta + \frac{1}{2\theta}\right)F.
\end{equation}
Therefore, $TP_{WKB}$ becomes,
\begin{equation}
\label{TPwkb}
    TP_{WKB} = \frac{|F|^2}{|A|^2} = \frac{4}{\left( 2\theta + \frac{1}{2\theta}\right)^2}. 
\end{equation}
For the present model, $\theta$ Eq. (\ref{theta0}) is given, with the aid of $V(a)$ Eq. (\ref{potential}) and $k(a)$ Eq. (\ref{ks}), by,
\begin{equation}
\label{theta}
    \theta = exp\left( 2\int_{a_l}^{a_r} \sqrt{\left( g_c a^2 - g_\Lambda a^4 - g_r - \frac{g_s}{a^2} - E \right)} da \right).
\end{equation}

\section{Results}
\label{results}

In the present section, we compute the $TP_{WKB}$ Eq. (\ref{TPwkb}), for the potential $V(a)$ Eq. (\ref{potential}) which has the shape of a well followed by a barrier. As discussed earlier, we consider the models where $g_c > 0$, $g_r \geq 0$, $g_\Lambda > 0$ and $g_s < 0$. In general, $TP_{WKB}$ is a function of $E$ and the HL parameters. In order to establish how $TP_{WKB}$ depends on those quantities, we fix all of them except the one we want to investigate. We repeat that procedure for many different values of all the quantities. In particular, when we study the dependence of $TP_{WKB}$ with the HL parameters, we fix the same energy value for all parameters. Another important quantities, we must determine in order to compute $TP_{WKB}$ are the scale factor values where the energy $E$ intercepts the potential $V(a)$ at the left ($a_l$) and at the right ($a_r$). In the following Subsections, we start that study.

\subsection{$g_\Lambda$}

\begin{figure}
	\centering
	\includegraphics[width=0.6\textwidth]{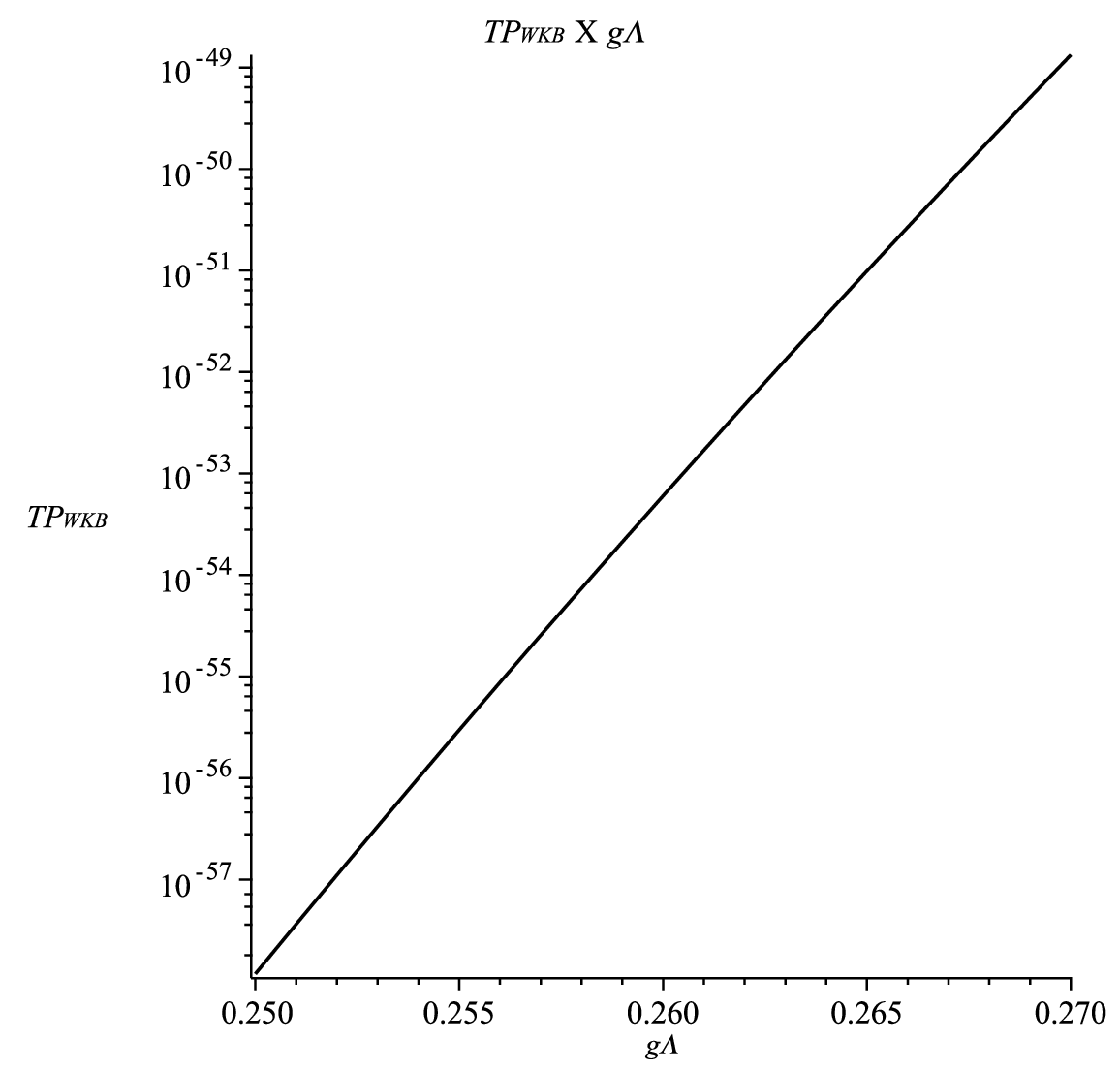}
	\caption{Variation of $TP_{WKB}$ as $g_{\Lambda}$ changes for a fixed energy $E = 100$.}
	\label{pot2-glambda}
\end{figure}

In order to study how $TP_{WKB}$ depends on the HL parameter $g_\Lambda$, we fix appropriate values of $E$, $g_c, g_r$, $g_s$ and varies $g_\Lambda$. We are considering only positive values of $g_\Lambda$, because they produce an accelerate expansion of the universe. We repeat that procedure for a great number of appropriate values of all those quantities and reach the following conclusion: $TP_{WKB}$ increases if one increases $g_\Lambda$. Therefore, it is more likely that the universe is born with the greatest possible value of that parameter. We show an example of that result for a model where $E=100$, $g_c=15, g_r=15$, and $g_s=-0.1$. In the present example, we choose $21$ different values of $g_\Lambda$: we start with $g_\Lambda = 0.250$ all the way up to $g_\Lambda = 0.270$ in steps of $0.001$. For all chosen values of $g_\Lambda$, the maxima values of the potentials $V(a)$ Eq. (\ref{potential}) are greater than $E=100$. 
In Figure \ref{pot2-glambda}, one can see the curve $TP_{WKB} \times g_\Lambda$.

\subsection{$g_c$}

\begin{figure}
	\centering
	\includegraphics[width=0.6\textwidth]{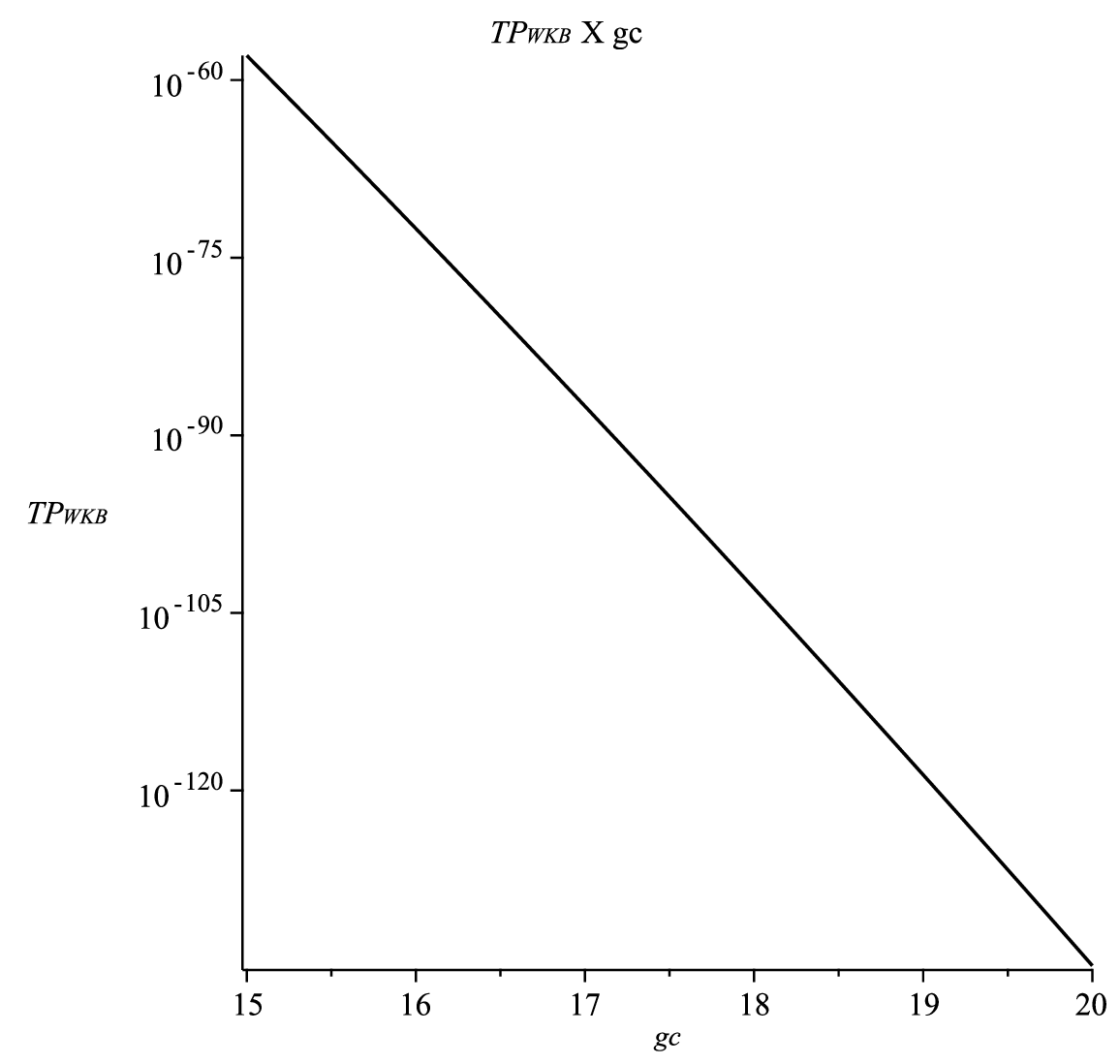}
	\caption{Variation of $TP_{WKB}$ as $g_{c}$ changes for a fixed energy $E = 100$.}
	\label{pot2-gc}
\end{figure}

In order to study how $TP_{WKB}$ depends on the HL parameter $g_c$, we fix appropriate values of $E$, $g_\Lambda, g_r$, $g_s$ and varies $g_c$. As we mentioned above, we are considering only positive values of $g_c$. We repeat that procedure for a great number of appropriate values of all those quantities and reach the following conclusion: $TP_{WKB}$ decreases if one increases $g_c$. Therefore, it is more likely that the universe is born with the smallest possible value of that parameter. We show an example of that result for a model where $E=100$, $g_\Lambda=0.25, g_r=15$, and $g_s=-0.1$. In the present example, we choose $26$ different values of $g_c$: we start with $g_c = 15.00$ all the way up to $g_c = 20.00$ in steps of $0.2$. For all chosen values of $g_c$, the maxima values of the potentials $V(a)$ Eq. (\ref{potential}) are greater than $E=100$. In Figure \ref{pot2-gc}, one can see the curve $TP_{WKB} \times g_c$.

\subsection{$g_s$}

\begin{figure}
	\centering
	\includegraphics[width=0.6\textwidth]{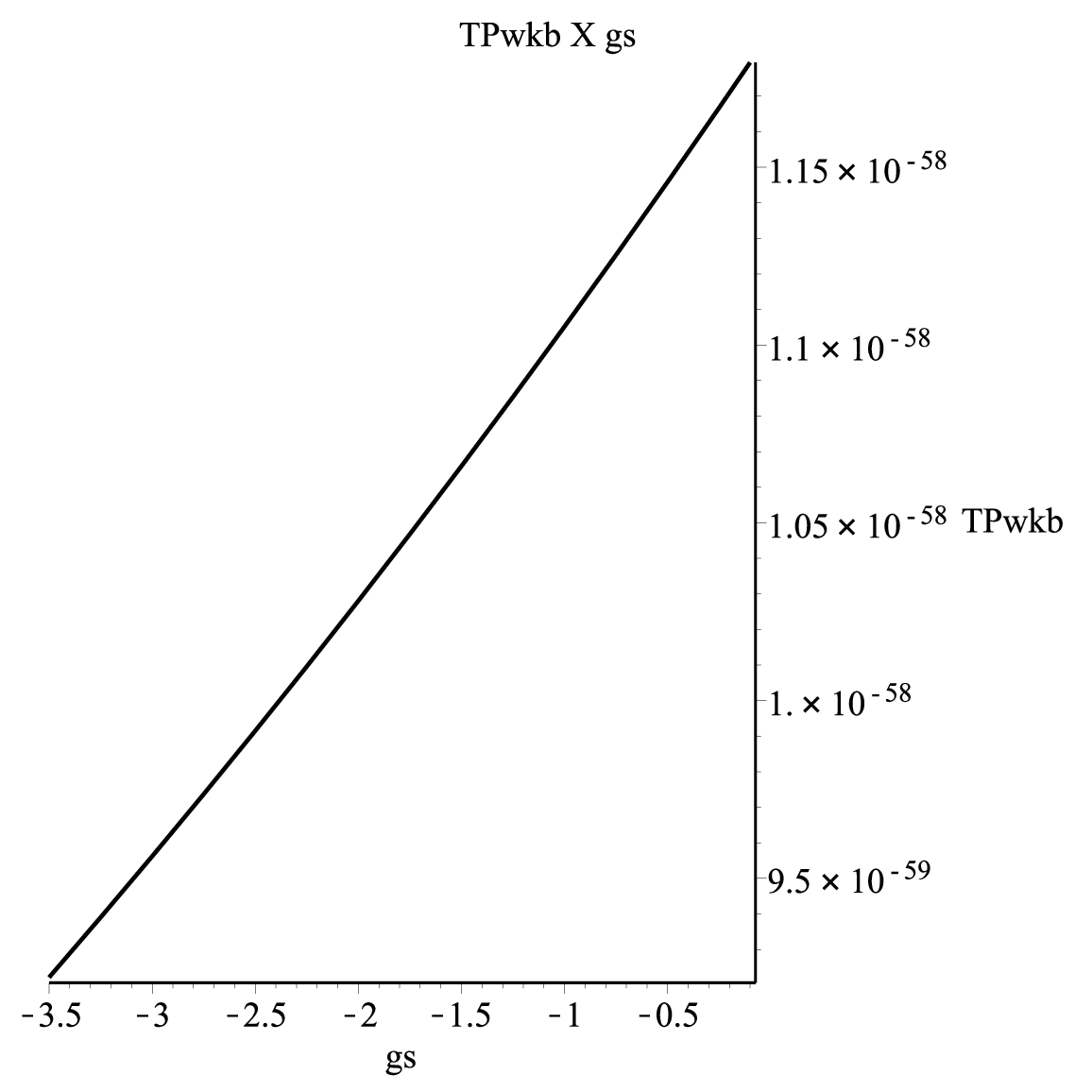}
	\caption{Variation of $TP_{WKB}$ as $g_{s}$ changes for a fixed energy $E = 100$.}
	\label{pot2-gs-neg}
\end{figure}

In order to study how $TP_{WKB}$ depends on the HL parameter $g_s$, we fix appropriate values of $E$, $g_\Lambda, g_r$, $g_c$ and varies $g_s$. As we mentioned above, we are considering only negative values of $g_s$. We repeat that procedure for a great number of appropriate values of all those quantities and reach the following conclusion: $TP_{WKB}$ increases if one increases $g_s$. Therefore, it is more likely that the universe is born with the greatest possible value of that parameter. We show an example of that result for a model where $E=100$, $g_\Lambda=0.25, g_r=15$, and $g_c=15$. In the present example, we choose $15$ different values of $g_s$: we start with $g_s = -0.10$, next $g_s = -0.25$ and then it goes all the way down to $g_s = -3.50$ in steps of $-0.25$. For all chosen values of $g_s$, the maxima values of the potentials $V(a)$ Eq. (\ref{potential}) are greater than $E=100$. In Figure \ref{pot2-gs-neg}, one can see the curve $TP_{WKB} \times g_s$.

\subsection{$g_r$}

\begin{figure}
	\centering
	\includegraphics[width=0.6\textwidth]{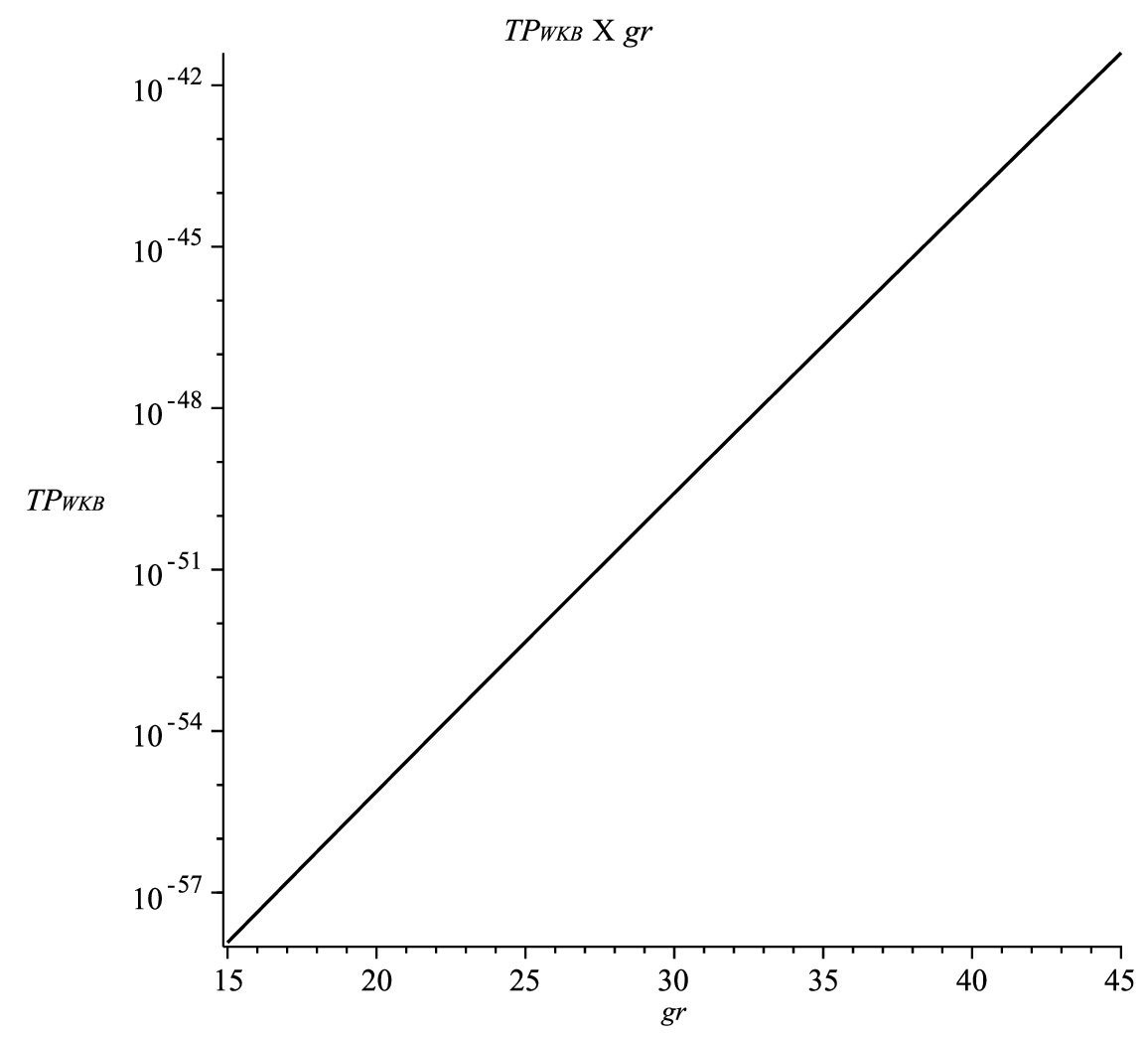}
	\caption{Variation of $TP_{WKB}$ as $g_{r}$ changes for a fixed energy $E = 100$.}
	\label{pot2-gr}
\end{figure}

In order to study how $TP_{WKB}$ depends on the HL parameter $g_r$, we fix appropriate values of $E$, $g_\Lambda, g_s$, $g_c$ and varies $g_r$. We are considering only positive values of $g_r$, because we want that the potential well has a negative minimum value. We repeat that procedure for a great number of appropriate values of all those quantities and reach the following conclusion: $TP_{WKB}$ increases if one increases $g_r$. Therefore, it is more likely that the universe is born with the greatest possible value of that parameter. We show an example of that result for a model where $E=100$, $g_\Lambda=0.25, g_s=-0.1$, and $g_c=15$. In the present example, we choose $21$ different values of $g_r$: we start with $g_r = 15.0$ all the way up to $g_r = 45.0$ in steps of $1.5$. For all chosen values of $g_r$, the maxima values of the potentials $V(a)$ Eq. (\ref{potential}) are greater than $E=100$. In Figure \ref{pot2-gr}, one can see the curve $TP_{WKB} \times g_r$.

\subsection{$E$}

\begin{figure}
	\centering
	\includegraphics[width=0.6\textwidth]{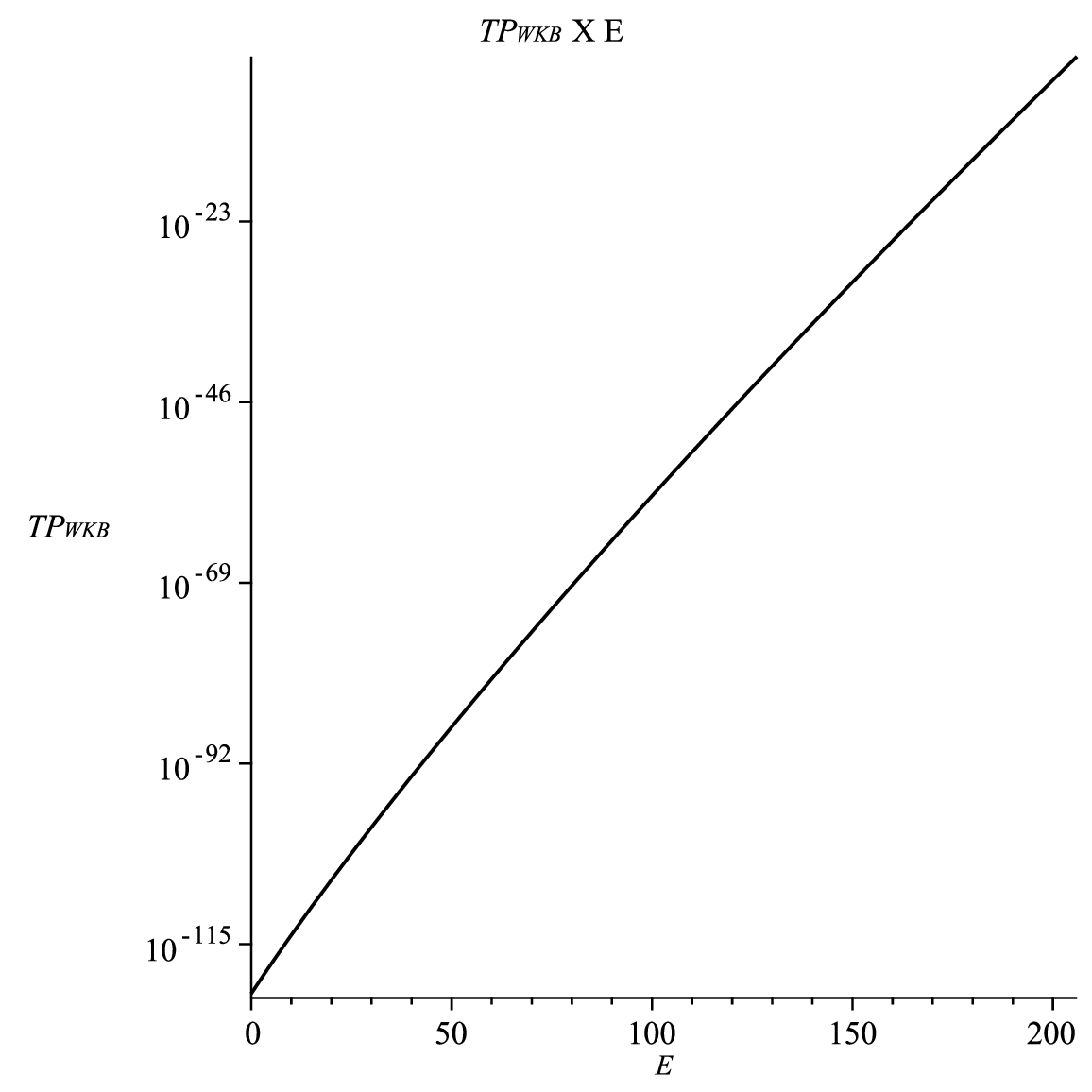}
	\caption{Variation of $TP_{WKB}$ as $E$ changes.}
	\label{pot2-em-neg}
\end{figure}

In order to study how $TP_{WKB}$ depends on the radiation fluid energy $E$, we fix appropriate values of $g_\Lambda, g_s$, $g_c$, $g_r$ and varies $E$. We are considering only positive values of $E$. We repeat that procedure for a great number of appropriate values of all those quantities and reach the following conclusion: $TP_{WKB}$ increases if one increases $E$. Therefore, it is more likely that the universe is born with the greatest possible value of the radiation fluid energy. We show an example of that result for a model where $g_r=15$, $g_\Lambda=0.25, g_s=-0.1$, and $g_c=15$. In the present example, we choose $55$ different values of $E$: we start with $E = 0$ all the way up to 
$E = 206$ in steps of different sizes. For all chosen values of $E$, the maximum value of the potential $V(a)$ Eq. (\ref{potential}) is greater than E. In Figure \ref{pot2-em-neg}, one can see the curve $TP_{WKB} \times E$.

\section{Conclusions}
\label{CC}

In the present paper, we investigated the birth of the universe in a cosmological model constructed 
using HL theory. We considered a homogeneous and isotropic FRW spacetime, with positively curved spatial sections ($k=1$).
The matter content of the model is a radiation perfect fluid. We restricted our attention to the projectable version of HL theory,
without the detailed balance condition. Initially, we wrote the total Hamiltonian of the gravitational sector of the model, with the
aid of the ADM formalism. Then, we obtained the total Hamiltonian of the matter sector of the model, with the aid of the Schutz variational
formalism \cite{schutz,schutz1}. Adding both total Hamiltonians, we found the total Hamiltonian of the model. From that total Hamiltonian, we
noticed that the potential barrier could have two different shapes. In the first one, it has a single barrier shape. In the second one, it has 
a well followed by a barrier shape. Due to the fact that the first barrier shape is a particular case of the second one, we restricted our 
attention to the investigation of the birth of the universe only for the second barrier shape. In order to learn more about that particular model, we studied the classical 
dynamics of the scale factor. First, using the total Hamiltonian of the model we drew a phase portrait of the model, which gave a qualitative 
idea on the scale factor behavior. Then, using once more the total Hamiltonian of the model we found and solved the Hamilton's equations. 
We quantitatively obtained all possible scale factor solutions for the model. After that, we quantized the model following Dirac's
formalism for constrained theories \cite{dirac,dirac4}, which produced the Wheeler-DeWitt equation. We solved that
equation using the WKB approximation and computed the appropriate tunneling probabilities for the birth of the universe ($TP_{WKB}$). Since $TP_{WKB}$ is a
function of the radiation energy and all the parameters coming from the HL theory, we investigated, 
separately, how it depends on all these quantities. As the result of those investigations, we determined that: $TP_{WKB}$ increases as
$E$, $g_\Lambda$, $g_s$ and $g_r$ increase; and it decreases as $g_c$ increases. Therefore, it is more likely that the universe is born with 
the greatest possible values of $E$, $g_\Lambda$, $g_s$ and $g_r$ and with the smallest possible value of $g_c$.

{\bf Acknowledgments}. A. Oliveira Castro Júnior thanks Coordena\c{c}\~{a}o de\\ Aperfei\c{c}oamento de Pessoal de N\'{i}vel Superior (CAPES).
G. A. Monerat thanks FAPERJ for financial support and Universidade do Estado do Rio de Janeiro (UERJ) for the Proci\^{e}ncia grant.


\appendix 

\section{Radiation fluid total Hamiltonian}
\label{SF}

In the Schutz variational formalism, for FRW cosmological models, the fluid four-velocity $U_{\nu}$ is given in terms of the following thermodynamical potentials $\mu$, $\phi, \theta$ and $S$,
\begin{equation}
\label{potencial}
U_{\nu} = \frac{1}{\mu}(\phi_{,\nu} +  \theta S_{,\nu}) \, ,
\end{equation}
where $\mu$ is the specific enthalpy, $S$ is the specific entropy and the potentials $\phi$ and $\theta$ have no clear physical meaning.
The four-velocity must satisfy the following normalization condition,
\begin{equation}
\label{normalization}
U^{\nu}U_{\nu} = -1
\end{equation}
In order to obtain the final result, we must consider few equations coming from thermodynamics, 
\begin{equation}  
\label{eqTermodinâmica}
   \rho = \rho_{0} (1 + \Pi),\qquad
   \mu = (1 + \Pi) + \frac{p}{\rho_{0}},\qquad
   TdS = d\Pi + pd\left(\frac{1}{\rho_{0}}\right),
\end{equation}
where $\Pi$ is the specific internal energy, $T$ is the absolute temperature and $\rho_{0}$ is the rest mass density. 
From the combination of equations (\ref{eqTermodinâmica}), it is possible to write,
\begin{equation} 
\label{eqTermodinâmica2}
   T = 1 + \Pi, \qquad
   S = \ln(1 + \Pi) \frac{1}{\rho_{0}^{\frac{1}{3}}}
\end{equation}
Next, one may express $\mu$ in terms of the thermodynamical potentials $S$, $\phi$ and $\theta$ Eq.(\ref{potencial}). In order to do that, one must use Eq.(\ref{normalization}),
\begin{equation} 
\label{entalpia}
    \mu = \frac{1}{N}(\dot{\phi} + \theta \dot{S}).
\end{equation}
In the next step, we want to write the radiation energy density $\rho$ as a function of $S$, $\phi$ and $\theta$. We do that with the aid of
Eqs. (\ref{eqTermodinâmica}), (\ref{eqTermodinâmica2}) and (\ref{entalpia}),
\begin{equation} 
\label{rho}
    \rho = \Bigg{(}\frac{\frac{1}{N}(\dot{\phi} + \theta \dot{S})}{\frac{4}{3}}\Bigg{)}^{4} e^{-3S}
\end{equation}
Now, we must introduce $\rho$ Eq.(\ref{rho}) in the perfect fluid action, which is given by,
\begin{equation} 
\label{pfaction}
S_f = \int d^4x \sqrt{-g} p,
\end{equation}
where $g$ is the determinant of the metric and $p$ is the fluid pressure. After doing that, we find with the aid of Eq.(\ref{eqstate}) the
following expression for the radiation fluid action,
\begin{equation} 
\label{Ação fluido Apêndice (2)}
\int_{M}d^{4}x\sqrt{-g} \frac{1}{3} \rho_{rad} = \int_{M}d^{4}x\sqrt{-g} \frac{1}{3}\Bigg{(}\frac{\frac{1}{N}(\dot{\phi} + \theta \dot{S})}{\frac{4}{3}}\Bigg{)}^{4} e^{-3S},
\end{equation}
It is straightforward to find the fluid Lagrangian from the action Eq.(\ref{Ação fluido Apêndice (2)}), which is given by,
\begin{equation} 
\label{lagrangiana fluido}
    L_{f} = \frac{27}{256} \frac{ a^{3}}{N^{3}} {(\dot{\phi} + \theta \dot{S})}^{4} e^{-3S}
\end{equation}
Using the usual definition, we can calculate the canonically conjugated momenta to the canonical variables $\phi$ ($p_{\phi}$) and $S$ ($p_{S}$), from the Lagrangian (\ref{lagrangiana fluido}).
We obtain the following values,
\begin{equation}  
\label{momentos canônicos fluido}
    p_{\phi} = \frac{\partial L_{f} }{\partial \dot{\phi}} = \frac{27}{64} \frac{ a^{3}}{N^{3}} {(\dot{\phi} + \theta \dot{S})}^{3} e^{-3S},
    \qquad
     p_{S} = \frac{\partial L_{f} }{\partial \dot{S}} = \theta   p_{\phi}
\end{equation}
Taking in account the canonical variables of the present model, the fluid total Hamiltonian $N \mathcal{H}_{f}$, is written in the following way,
\begin{equation} 
\label{hamiltoniana fluido}
N \mathcal{H}_{f} = \dot{\phi} p_{\phi} + \dot{S} p_{S} - N L_{f},
\end{equation}
Now, introducing $L_{f}$ Eq.(\ref{lagrangiana fluido}) and the momenta $p_{\phi}$ and $p_{S}$ Eq.(\ref{momentos canônicos fluido}) in $N \mathcal{H}_{f}$ Eq.(\ref{hamiltoniana fluido}), we obtain,
\begin{equation} 
\label{hamiltoniana fluido_1}
   N \mathcal{H}_{f} = \frac{ {p_{\phi}}^{\frac{4}{3}}}{a} e^{S}.
\end{equation}
It is possible to simplify the expression of $N \mathcal{H}_{f}$  Eq.(\ref{hamiltoniana fluido_1}) with the aid of the canonical transformations \cite{rubakov1,nivaldo},
\begin{equation} 
\label{transformação canônica}
    T = p_{s}e^{-S}{p_{\phi}}^{-\frac{4}{3}},
		\qquad
    p_{T} = {p_{\phi}}^{\frac{4}{3}}e^{S},
		\qquad
    \bar{\phi} = \phi - \frac{4}{3} \frac{p_{S}}{p_{\phi}},
		\qquad
    \bar{p}_{\phi} = p_{\phi}.    
\end{equation}
Finally, after writing $N \mathcal{H}_{f}$ Eq.(\ref{hamiltoniana fluido_1}) in terms of the new canonical variables and their conjugated momenta Eqs.(\ref{transformação canônica}), we obtain
the final expression for the radiation fluid total Hamiltonian,
\begin{equation} 
\label{hamiltoniana fluido_2}
   N\mathcal{H}_{f} = \frac{p_{T}}{a}.
\end{equation}
From that equation (\ref{hamiltoniana fluido_2}), it is not difficult to see that $T$, shall play the role of time in the quantum version of the present models.



\begin{thebibliography}{99}

\bibitem{horava} P. Ho{\v r}ava, {\it Quantum gravity at a Lifshitz point}, Phys. Rev. D {\bf 79}, 084008 (2009).

\bibitem{misner} R. Arnowitt, S. Deser and C. W. Misner, {\it The Dynamics of General Relativity}, in {\it Gravitation: an introduction to current 
research}, ed. L. Witten (Wiley, New York, 1962), Chapter 7, pp 227-264 and arXiv:gr-qc/0405109.

\bibitem{mattvisser1} T. P. Sotiriou, M. Visser and S. Weinfurtner, {\it Quantum gravity without Lorentz invariance}, JHEP {\bf 10},
033 (2009).

\bibitem{wang} A. Wang and R. Maartens, {\it Linear perturbation of cosmological models in the Ho\v{r}ava-Lifshitz theory of gravity without detailed balance}, Phys. Rev. D {\bf 81}, 024009 (2010).

\bibitem{bertolami} O. Bertolami and C. A. D. Zarro, {\it Ho\v{r}ava-Lifshitz quantum cosmology}, Phys. Rev. D {\bf 84}, 044042 (2011).

\bibitem{saa} J. P. M. Pitelli and A. Saa, {\it Quantum singularities in Ho\v{r}ava-Lifshitz cosmology}, Phys. Rev. D {\bf 86}, 063506 (2012).

\bibitem{kord} B. Vakili and V. Kord, {\it Classical and quantum Ho\v{r}ava-Lifshitz cosmology in a minisuperspace perspective}, Gen. Relativ. Gravit. {\bf 45}, 1313 (2013).

\bibitem{pedram2} H. Ardehali and P. Pedram, {\it Chaplygin gas Ho\v{r}ava-Lifshitz quantum cosmology}, Phys. Rev. D {\bf 93}, 043532 (2016).

\bibitem{misonoh} Y. Misonoh, M. Fukushima and S. Miyashita, {\it Stability of singularity-free cosmological solutions in Ho\v{r}ava-Lifshitz gravity}, Phys. Rev. D {\bf 95}, 044044 (2017).

\bibitem{ivano} R. Maier and I. D. Soares, {\it Ho\v{r}ava-Lifshitz bouncing Bianchi IX Universes: A dynamical system analysis}, Phys. Rev. D {\bf 96}, 103532 (2017).

\bibitem{bramberger} S. F. Bramberger et al, {\it Solving the flatness problem with an anisotropic instanton in Ho\v{r}ava-Lifshitz gravity}, Phys. Rev. D {\bf 97}, 043512 (2018).

\bibitem{gil3} G. Oliveira-Neto, L. G. Martins, G. A. Monerat and E. V. Corr\^{e}a Silva, {\it De Broglie-Bohm interpretation of a Ho\v{r}ava-Lifshitz quantum cosmology model}, Mod. Phys. Lett. A {\bf 33}, 1850014 (2018).

\bibitem{gil} G. Oliveira-Neto, L. G. Martins, G. A. Monerat and E. V. Corr\^{e}a Silva, {\it Quantum cosmology of a Ho\v{r}ava-Lifshitz model coupled to radiation}, Int. J. Mod. Phys. D {\bf 28}, 1950130 (2019).

\bibitem{gil2} E. M. C. Abreu, A. C. R. Mendes, G. Oliveira-Neto, J. Ananias Neto, L. G. Rezende Rodrigues and M. Silva de Oliveira, {\it 
Ho\v{r}ava–Lifshitz cosmological models with noncommutative phase space variables}, Gen. Relativ. Gravit. {\bf 51}, 95 (2019).

\bibitem{gao} F. Gao and J. Llibre, {\it Global dynamics of the Ho\v{r}ava-Lifshitz cosmological system}, Gen. Relativ. Gravit. {\bf 51}, 152 (2019).

\bibitem{leon} G. Leon, A. Paliathanasis, {\it Extended phase-space analysis of the Ho\v{r}ava-Lifshitz cosmology}, Eur. Phys. J. C {\bf 79}, 746 (2019).

\bibitem{cordero} R. Cordero, H. G. Compe\'{a}n and F. J. Turrubiates, {\it A phase space description of the FLRW quantum cosmology
in Ho\v{r}ava-Lifshitz type gravity}, Gen. Relativ. Gravit. {\bf 51}, 138 (2019).

\bibitem{nilsson} N. A. Nilsson and E. Czuchry, {\it Ho\v{r}ava-Lifshitz cosmology in light of new data}, Phys. Dark Univ. {\bf 23}, 100253 (2019).

\bibitem{gao1} F. Gao and J. Llibre, {\it Global dynamics of Ho\v{r}ava-Lifshitz cosmology with non-zero
curvature and a wide range of potentials}, Eur. Phys. J. C {\bf 80}, 137 (2020).

\bibitem{compean} H. G. Compe\'{a}n and A. V\'{a}zquez, {\it Euclidean wormholes in Ho\v{r}ava-Lifshitz gravity}, Phys. Rev. D {\bf 101}, 084048 (2020).

\bibitem{gao2} F. Gao and J. Llibre, {\it Global Dynamics of the Ho\v{r}ava-Lifshitz Cosmological Model
in a Non-Flat Universe with Non-Zero Cosmological Constant}, Universe {\bf 7}, 445 (2021).

\bibitem{tavakoli} F. Tavakoli, B. Vakili and H. Ardehali, {\it Ho\v{r}ava-Lifshitz Scalar Field Cosmology: Classical and
Quantum Viewpoints}, Adv. High Energy Phys. 2021, 6617910 (2021).

\bibitem{vicente} G. S. Vicente, {\it Quantum Ho\v{r}ava-Lifshitz cosmology in the de Broglie–Bohm interpretation}, Phys. Rev. D {\bf 104}, 103525 (2021).

\bibitem{gao3} F. Gao and J. Llibre, {\it Global dynamics of the Hořava–Lifshitz cosmology in the presence of
non-zero cosmological constant in a flat space}, Phys. Dark Univ. {\bf 38}, 101139 (2022).

\bibitem{compean1} H. G. Compe\'{a}n and D. M. Pacheco, {\it Lorentzian Vacuum Transitions in Hořava–Lifshitz Gravity}, Universe {\bf 8}, 237 (2022).

\bibitem{compean2} H. G. Compe\'{a}n and D. M. Pacheco, {\it Generalized uncertainty principle effects in the Hořava–Lifshitz quantum theory of gravity}, Nucl. Phys. B {\bf 977}, 115745 (2022).

\bibitem{czuchry} E. Czuchry, {\it Resolution of Cosmological Singularity in Hořava–Lifshitz Cosmology}, Universe {\bf 9}, 160 (2023).

\bibitem{nilsson1} E. Di Valentino, N. A. Nilsson and M. I. Park, {\it A new test of dynamical dark energy models and cosmic tensions in Ho\v{r}ava-Lifshitz gravity}, Mon. Not. Roy. Astron. Soc. {\bf 519}, 5043 (2023).

\bibitem{wang1} A. Wang, {\it Ho\v{r}ava gravity at a Lifshitz point: A progress report}, Int. J. Mod. Phys. D {\bf 26}, 1730014 (2017).

\bibitem{dirac} P. A. M. Dirac, {\it Generalized Hamiltonian dynamics}, Can. J. Math. \textbf{2}, 129 (1950).

\bibitem{dirac1} P. A. M. Dirac, {\it Generalized Hamiltonian dynamics}, Proc. Roy. Soc. London A \textbf{249}, 326 (1958).

\bibitem{dirac2} P. A. M. Dirac, {\it The Theory of gravitation in Hamiltonian form}, Proc. Roy. Soc. London A \textbf{249}, 333 (1958).

\bibitem{dirac3} P. A. M. Dirac, {\it Fixation of coordinates in the Hamiltonian theory of gravitation}, Phys. Rev. \textbf{114}, 924 (1959).

\bibitem{dirac4} P. A. M. Dirac, {\it Lectures on Quantum Mechanics}, Belfer Graduate School of Science Monographs Series, Number 2, 
(Yeshiva University, New York, 1964).

\bibitem{dewitt} B. S. DeWitt, {\it Quantum Theory of Gravity. 1. The Canonical Theory}, Phys. Rev. D \textbf{160}, 1113 (1967).

\bibitem{wheeler} J. A. Wheeler, {\it Superspace and the nature of quantum geometrodynamics}, in {\it Batelles Rencontres}, eds. C. DeWitt
and J. A. Wheeler (Benjamin, New York, 1968), 242.

\bibitem{halliwell} J. J. Halliwell, {\it Introductory Lectures on Quantum Cosmology}, in \emph{Quantum Cosmology and Baby Universes}, 
Jerusalem Winter School for Theoretical Physics vol. 7, eds. by S. Coleman, J. B. Hartle, T. Piran and S. Weinberg 
(World Scientific, Singapore, 1991).

\bibitem{paulo} P. Vargas Moniz, {\it Quantum Cosmology - The Supersymmetric Perspective - vol. 1: Fundamentals}, 
Lect. Notes Phys. 803 (Springer, Berlin Heidelberg, 2010).

\bibitem{kiefer} C. Kiefer, {\it Quantum Gravity} (3rd edition), (Oxford University Press, Oxford, 2012).

\bibitem{julio} N. Pinto-Neto and J. C. Fabris, {\it Quantum cosmology from the de Broglie Bohm
perspective}, Class. Quantum Grav. {\bf 30}, 143001 (2013).

\bibitem{grishchuk} L. P. Grishchuk and Ya. B. Zeldovich, in {\it Quantum Structure of Space and Time}, eds. M. Duff and C. Isham 
(Cambridge University Press, Cambridge, 1982).

\bibitem{vilenkin} A. Vilenkin, {\it Creation of Universes from Nothing}, Phys. Lett. B \textbf{117}, 25 (1982). 

\bibitem{vilenkin1} A. Vilenkin, {\it Quantum Creation of Universes}, Phys. Rev. D \textbf{30}, 509 (1984).

\bibitem{vilenkin2} A. Vilenkin, {\it Boundary Conditions in Quantum Cosmology}, Phys. Rev. D \textbf{33}, 3560 (1986).

\bibitem{hawking} J. B. Hartle and S. W. Hawking, {\it Wave Function of the Universe}, Phys. Rev. D \textbf{28},
2960 (1983).

\bibitem{linde} A. D. Linde, {\it Quantum Creation of the Inflationary Universe}, Lett. Nuovo Cim. \textbf{39}, 401 (1984).

\bibitem{rubakov} V. A. Rubakov, {\it Quantum Mechanics in the Tunneling Universe}, Phys. Lett. B \textbf{148}, 280 (1984).

\bibitem{vilenkin3} For a critical review see: A. Vilenkin, {\it Quantum cosmology and eternal inflation}, in
{\it Cambridge 2002, The future of theoretical physics and cosmology}, 
eds. G. W. Gibbons, E. P. S. Shellard and S. J. Rankin (Cambridge University
Press, Cambridge, 2003), 649-666.

\bibitem{paulo1} Mariam Bouhmadi-Lopez and Paulo Vargas Moniz, {\it FRW quantum cosmology with a generalized Chaplygin gas}, 
Phys. Rev. D \textbf{71}, 063521 (2005).

\bibitem{acacio} J. Acacio de Barros, E. V. Corr\^{e}a Silva, G. A. Monerat, G. Oliveira-Neto, L. G. 
Ferreira Filho and P. Romildo Jr, {\it Tunneling probability for the birth of an asymptotically 
DeSitter universe}, Phys. Rev. D {\bf 75}, 104004 (2007), [arXiv:0612031 [gr-qc]].

\bibitem{germano} G. A. Monerat, G. Oliveira-Neto, E. V. Corr\^{e}a Silva, L. G. Ferreira Filho, P. Romildo Jr., J. C. Fabris, 
R. Fracalossi, S. V. B. Gon\c{c}alves and F. G. Alvarenga, {\it The dynamics of the early universe and the initial conditions 
for inflation in a model with radiation and a Chaplygin gas}, Phys. Rev. D {\textbf 76}, 024017 (2007), [arXiv:0704.2585 [gr-qc]].

\bibitem{germano1} G.A. Monerat, C.G.M. Santos, G. Oliveira-Neto, E.V. Corr\^{e}a Silva and L. G. Ferreira Filho, {\it The dynamics 
of the early universe in a model with radiation and a generalized Chaplygin gas}, Eur. Phys. J. Plus {\bf 136}, 34 (2021).

\bibitem{germano2} G.A. Monerat, F.G. Alvarenga, S.V.B. Gon\c{c}alves, G. Oliveira-Neto, C.G.M. Santos, E.V. Corr\^{e}a Silva, {\it
The effects of dark energy on the early Universe with radiation and an ad hoc potential}, Eur. Phys. J. Plus {\bf 137}, 117 (2022).

\bibitem{rocha} N. M. N da Rocha, G. A. Monerat, F. G. Alvarenga, S. V. B. Gon\c{c}alves, G. Oliveira-Neto, E. V. Corr\^{e}a Silva, 
C. G. M. Santos, {\it  Early universe with dust and Chaplygin gas}, Eur. Phys. J. Plus {\bf 137}, 1103 (2022).

\bibitem{gil4} G. Oliveira-Neto, D. L. Canedo, G. A. Monerat, {\it Tunneling probabilities for the birth of universes with radiation, 
cosmological constant and an ad hoc potential}, Eur. Phys. J. Plus {\it 138}, 400 (2023). 

\bibitem{schutz} B. F. Schutz, {\it Perfect Fluids in General Relativity: Velocity Potentials and a Variational Principle}, 
Phys. Rev. D \textbf{2}, 2762, (1970).

\bibitem{schutz1} B. F. Schutz, {\it Hamiltonian Theory of a Relativistic Perfect Fluid}, Phys. Rev. D \textbf{4}, 3559, (1971).

\bibitem{merzbacher}  E. Merzbacher, \emph{Quantum Mechanics}. 3rd ed. (John Wiley \& Sons, Inc., New York, 1998), Chap. 7.

\bibitem{griffiths} D. J. Griffiths, \emph{Introduction to Quantum Mechanics}. 2nd ed. (Prentice Hall, New Jersey, 2005), Chap. 8.

\bibitem{rubakov1} V. G. Lapchinskii and V. A. Rubakov, {\it Quantum gravitation: Quantization of the Friedmann model}, 
Theor. Math. Phys. {\bf 33}, 1076 (1977).

\bibitem{nivaldo} F. G. Alvarenga, J.C. Fabris, N. A. Lemos and G. A. Monerat, {\it Quantum Cosmological Perfect Fluid Models}, 
Gen. Rel. Grav. {\bf 34}, 651 (2002).


\end{thebibliography}
\end{document}